\documentclass[twocolumn,showpacs,preprintnumbers,amssymb]{revtex4-1}

\usepackage{graphicx}
\usepackage{dcolumn}
\usepackage{color}
\usepackage{bm}

\def \be{\begin{equation}}
\def \ee{\end{equation}}
\def \ben{\begin{eqnarray}}
\def \een{\end{eqnarray}}

\begin{document}

\title{Multi-wave-mixing-induced nonlinear effects in an electromagnetically induced grating}

\author{Bibhas Kumar Dutta}
\altaffiliation{bibhas\_k\_dutta@yahoo.co.in}
\affiliation{Department of Physics, Sree Chaitanya College (WB State University), Habra, North 24 Parganas - 743 268, W. B., India}

\author{Pradipta Panchadhyayee}
\altaffiliation{ppcontai@gmail.com}
\affiliation{Department of Physics (UG \& PG), Prabhat Kumar College (Vidyasagar University), Contai, Purba Medinipur-721404, India}

\author{Indranil Bayal}
\altaffiliation{indranil\_bayal@yahoo.com}
\affiliation{ITER, Siksha ’O’ Anusandhan University, Bhubaneswar-751030, Odisha, India}

\author{Nityananda Das}
\altaffiliation{ndas228@yahoo.com}
\affiliation{Department of Physics, J. K. College (Sidho Kanho Birsha University)
Purulia - 723 101, W. B., India}

\author{Prasanta Kumar Mahapatra}
\altaffiliation{mahapatrap@yahoo.com}
\affiliation{ITER, Siksha ’O’ Anusandhan University, Bhubaneswar-751030, Odisha, India}

\begin{abstract}
We propose a multi-field-coupled atomic model that exhibits controllable $symmetric$ and $asymmetric$
evolution of significantly enhanced diffraction peaks in an opto-atomic
grating at far-field regime.
Such results are obtained by the linear and nonlinear modulation of
the intensities of the diffraction peaks as a result of multi-wave-mixing-induced 
modification of spatially modulated coherence in a closed
four-level atomic system. Novelty of the results lies in predicting super
symmetric alignment of the diffraction peaks due to the dominance of the
amplitude part of the grating-transfer-function at the condition of
exact atom-field resonance, which is unique to the present model. Efficacy
of the present scheme is to apply it in producing nonlinear light generated by four-wave-mixing-induced control of spatially modulated coherence effect. The work also finds its importance for its applicability in the field of high-precision atomic lithography.
\end{abstract}



\maketitle

\section{Introduction}

With the advancement of laser based science and technology, it has been
possible to construct new optical device like Electromagnetically Induced
Grating (EIG) [1-3].
In the case of Electromagnetically Induced Diffraction (EID),
any aperture, or obstacle responsible for diffracting light beam can be built directly
or indirectly by one or more external electromagnetic fields [1].
Consequently, the formation of a grating as a result of the way of applying
the external fields interacting resonantly with an atomic medium gives rise
to the concept of electromagnetically induced grating (EIG). 
In contrast to the four-wave-mixing (FWM) induced
grating [1-3], the features of spatially modulated absorption and transparency
of a laser beam passing through a coherently prepared atomic medium are explored to 
demonstrate EIG [4-8] both theoretically and experimentally at
far-field regime, where $z$$>>$$D^{2}/{\lambda}$; $z$ being the distance of probing 
the output between the grating and the image plane, $D$, the lateral
dimension of the aperture and $\lambda$, the wavelength of the incident wave.

We note that the phenomena like Electromagnetically Induced Transparency
(EIT)[9], or double-EIT (DEIT)[10], or Electromagnetically Induced Absorption
(EIA)[11], laser-controlled Decay Interference Induced Coherence (DIIC) [12]
may lead to obtain $spatially$ $modulated$ $atomic$ $coherence$ under the
$standing$-$wave$ $field$ configuration. Owing to such coherence
effects, a number of works [4-8,13-30] have presented EIG in various atomic
models. More specifically, EIG in terms of spatially modulated EIT or DEIT
has been analyzed in ref.[4-8,13,14,18,20,22-24] and EIG based on EIA has been
explained in ref.[5,17]. Impact of the DIIC effect on the spatially modulated coherence is
reported in ref.[15,16,19] and the gain-assisted control of EIG has been shown in
ref.[21]. Contribution of coherent phase-modulation of the transfer function,
which is attributed due to the presence of standing-wave regime, has been shown to be
essential for the enhancement of the first and second order peaks when compared
to the central peak [4,7,8,15-20]. The appearance of EIG with Rydberg atoms has also been discussed
in ref.[23]. Similar features associated with EIG in generic models are explored in solid state media such as, semiconductor quantum wells [26-29].
Evolution of optical ${\cal {PT}}$ -symmetry by asymmetric diffraction in
coherence controlled Raman-Nath grating has been investigated in ref.[30].
Cross-grating like structures based on two-dimensional (2D) EIG is described in
ref.[31-33]. The mechanism of electromagnetically induced Talbot effect
similar to EIG has been discussed in ref.[34] and also demonstrated
experimentally [35].

\begin{figure*}
\begin{center}
\includegraphics[width=0.5\linewidth]{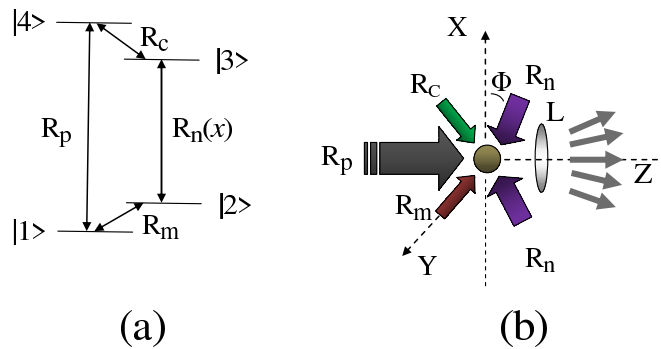}
\end{center}
\caption
{(a) Field-coupled energy level diagram of a four-level atomic system.
         $R_j$ ($j = p, m, n, c$) denotes the Rabi frequency for the applied
         fields. Specifically, $R_n(x)$ denotes the position dependence
         of the Rabi frequency (see text).
      (b) Schematic view of possible field arrangement with the atom
         (circular spot) placed at the centre. Angle $\phi$ denotes
         the orientation of the components of the control fields forming
         standing wave. $L$ indicates a lens. Diverging arrows
         are shown for different orders of diffraction.} 
\end{figure*}   

The mechanism of EID rendering EIG as presented in this article may lead
to construct a device called to be $opto-atomic$ $grating$ for the study of
diffraction of laser at high frequency region. The scheme of EIG can be
chosen to be an useful technique to generate nonlinear light almost
free from background-noise as reported in ref.[2,3] for detection of light
signal based on the four-wave-mixing (FWM) technique. In this article,
electromagnetically induced far-field diffraction of a weak
coherent optical beam is investigated after passing it through a atomic
system interacting with the assembly of travelling and standing wave fields
as shown in Fig.1(a,b). Fig.1(a) represents the field-coupled energy level
diagram for a close loop interaction scheme, where, in general, the linear
response of a weak probe field is modified by the nonlinear effect generated
by the $multi$-$wave$-$mixing$-$induced$ $coherence$ (MWMIC). The later one is found to
have significant contribution to adjust the opto-atomic slits originated by
spatially modulated coherence in the standing-wave regime.

In the proposed scheme, generation of spatially modulated coherence is
regulated not only by the DEIT, but by the MWMIC also. Such controlling
process, at the condition of exact atom-field resonance, is shown to
be unique to the given four-level model in inducing supersymmetric structure in 
the occurrence of diffraction pattern where peaks of all the orders have almost equal magnitude 
with significant sharpness. This is worth mentioning that the amplitude part of the
grating-transfer-function plays the vital role in forming such peak pattern.
Attempt has been made to exhibit a number of sharp peaks to be
appearing in the output of the grating. It has been shown that, for small
shifting from the condition of exact resonance, nonlinear peak modulation
resulted from the effect of MWMIC leads us to obtain controllable
asymmetry of the alignment of the higher order peaks.
We have examined the explicit role of the Rabi frequencies and detuning on
the intensities of the higher order peaks. The present model seems to be an useful technique for generation and detection of nonlinear light as a result of four-wave-mixing induced control of spatially modulated coherence effect. Overall, the diffraction patterns as obtained in this work seem to be plausible in practice for optically-induced atomic lithography.   

\section{Theoretical Model}

The field-coupled energy-level configuration of the atom is shown in Fig.1(a),
where the transitions $|1>$-$|2>$, $|1>$-$|4>$, $|2>$-$|3>$ and $|3>$-$|4>$
are influenced by the field induced Rabi frequencies
$R_m$$=$$\frac{{\bar\mu}_{21}.{\bar{\epsilon}}_m}{2\hbar}$,
$R_p$$=$$\frac{{\bar\mu}_{41}.{\bar{\epsilon}}_p}{2\hbar}$,
$R_n(x)$$=$$R_n$$sin({\pi}px/{{\Lambda}_c})$ with
$R_n$$=$$\frac{{\bar\mu}_{32}.{\bar{\epsilon}}_n}{2\hbar}$, and
$R_c$$=$$\frac{{\bar\mu}_{43}.{\bar{\epsilon}}_c}{2\hbar}$ respectively.
Here, $\mu_{jk}$ stands for the corresponding transition moment, while
${\epsilon_j}/2$ for the field amplitude. The field probing the
coherence induced by the control fields in the system is designated by
the Rabi frequency $R_p$. Spatial dependence of the Rabi
frequency denoted as $R_n(x)$ originates from the standing wave formed by
the counter-propagating components of the coupling field (Fig.1(b)) defined
as $E_n(x,t)$$=$$\frac{\bar{\epsilon}_n}{2}$$sin(k_c x cos\phi)$
$e^{i\omega_nt}$ + $c.c.$$=$
$\frac{\bar{\epsilon}_n}{2}$$sin({\pi}px/{{\Lambda}_c})$
$e^{i\omega_nt}$ + $c.c.$; where $p$$=$$cos{\phi}$,
and
${\Lambda}_c$$=$${{\lambda}_c}/2$
 with ${\lambda}_c$ being the wavelength of the coupling field.
We note that
${\Lambda}^{\prime}_c$$(=$${{\Lambda}_c}/p)$ implies the separation between two consecutive nodes,
or antinodes. By changing the angle $\phi$ the value of $\Lambda^{\prime}_c$
can be varied. Other control fields of Rabi frequencies $R_m$ and $R_c$
are treated to be the travelling waves like the probe field in the present
model.

In view of finding the validity of the given model in a realistic atomic system, we have chosen the field-induced transitions for Rubidium 87$D_1$ lines
($5^2S_{1/2}$ $\leftrightarrow$ $5^2P_{1/2}$). The dipole-allowed transitions $|1>$-$|4>$ and $|2>$-$|3>$ coincide with $(S)F = 1$ $\leftrightarrow$ $(P)F = 2$ and $(S)F = 2$ $\leftrightarrow$ $(P)F = 1$ transitions. Dipole forbidden transitions $|1>$-$|2>$ and $|3>$-$|4>$ correspond to $(S)F = 1$ $\leftrightarrow$ $(S)F = 2$ and $(P)F = 1$ $\leftrightarrow$ $(P)F = 2$ are taken into account in the presence of microwave field.  

The coherent part of the atom-field interaction is described by the
Hamiltonian under the electric dipole and the rotating wave approximations as
\begin{widetext}
\be
{\cal H}(t) = -{\hbar}[\Delta_m |2><2| + (\Delta_p - \Delta_c) |3><3| + \Delta_p |4><4| 
            + (R_m |1><2|  + R_p |1><4| \\
            + R_n(x) |2><3|  + R_c |3><4| + c.c)]           
\ee
\end{widetext}

where $\Delta_m$$=$$\Delta_p - \Delta_c - \Delta_n$ with the detuning
parameters $\Delta_p$$=$$\omega_{p} - \omega_{41}$,
$\Delta_c$$=$$\omega_{c} - \omega_{43}$,
$\Delta_n$$=$$\omega_{n} - \omega_{32}$ and
$\Delta_m$$=$$\omega_{m} - \omega_{21}$. The system dynamics can be
explained by the semiclassical density matrix equation as given by
\be
\frac{\partial {\rho}}{\partial t} = -\frac{i}{\hbar}[{\cal H},\rho]
                                        + \Lambda \rho
\ee
where the term $\Lambda \rho$ [14] includes the effect of incoherent
decay-mechanism inherent to the present atomic model. The required
off-diagonal density matrix equations are presented as follows
\be
\dot{\rho}_{41} = -Z_{41} {\rho}_{41}
                     + iR^*_p (\rho_{11} - \rho_{44}) + iR^*_c \rho_{31}
                     -iR^*_m \rho_{42}
\ee
\be
\dot{\rho}_{31} = -Z_{31} {\rho}_{31}
                    + iR^*_n(x) \rho_{21} + iR_c \rho_{41}
                    - iR^*_m \rho_{32} - iR^*_p \rho_{34}
\ee
\be
\dot{\rho}_{21} = -Z_{21} {\rho}_{21}
                    + iR^*_m (\rho_{11} - \rho_{22}) + iR_n(x) \rho_{31}
                    - iR^*_p \rho_{24} 
\ee
where $Z_{41}$$=$$(\gamma_{41} + \gamma_{42})/2 - i\Delta_p$,
$Z_{31}$$=$$(\gamma_{31} + \gamma_{32})/2 - i(\Delta_p - \Delta_c)$ and
$Z_{21}$$=$$\Gamma_{21} - i(\Delta_p - \Delta_c - \Delta_n)$. Here,
$\gamma_{mn}$ ($m=3,4$ and $n=1,2$) denotes the natural decay rate, while
$\Gamma_{21}$ is considered to incorporate small coherence dephasing rate.

Under weak-field approximation, we treat the Rabi frequencies $R_p$
and $R_m$ to the first order and the others ($R_n$ and $R_c$) to all orders,
the given set of density matrix equations can be solved in steady state to
obtain the expression of ${\rho}^{(1)}_{41}$ on the basis of following conditions
to be satisfied: $\rho^{(0)}_{11}$$\approx$$1$, $\rho^{(0)}_{42}$$=$
$\rho^{(0)}_{32}$$=$$\rho^{(0)}_{34}$$=$$0$, and $\rho_{jk}$$=$
$\rho^*_{kj}$. By making the substitutions: $\rho^{(1)}_{41}$$=$
$\tilde{\rho}_{41}$$e^{-i\phi_p}$, $R_p$$=$$|R_p|$$e^{i\phi_p}$,
$R_m$$=$$|R_m|$$e^{i\phi_m}$, $R_n(x)$$=$$|R_n(x)|$$e^{i\phi_n}$,
$R_c$$=$$|R_c|$$e^{i\phi_c}$ with the
condition $\phi_p$$=$$\phi_m + \phi_n + \phi_c$, we obtain
\be
\frac{\tilde{\rho}_{41}}{|R_p|} = f_L + f_{NL1} + f_{NL2}
\ee
with
$$
f_L = i\frac{Z_{21}Z_{31}}
        {Z_{21}Z_{31}Z_{41} + Z_{21}|R_c|^2 + Z_{41}|R_n(x)|^2},
$$
$$
f_{NL1} = i\frac{|R_n(x)|^2}
            {Z_{21}Z_{31}Z_{41} + Z_{21}|R_c|^2 + Z_{41}|R_n(x)|^2},
$$
$$
f_{NL2} = -i\frac{|R_n(x)||R_m||R_c|/|R_p|}
            {Z_{21}Z_{31}Z_{41} + Z_{21}|R_c|^2 + Z_{41}|R_n(x)|^2},
$$
where $f_L$ indicates the linear response of the probe field,
while the nonlinear terms $f_{NL1}$ and $f_{NL2}$ are for inducing
cross-phase modulation and MWMIC in the probe response. The polarization
induced in the probe transition is given by performing the quantum average
over the corresponding transition moment [12,14] as follows
\be
{\cal P}_p = \epsilon_0 \chi_p \epsilon_p
            = 2N{\mu}_{14}\tilde{\rho}_{41}
\ee
where $\epsilon_0$ being the free-space permittivity and $N$, the atomic
density. The susceptibility $\chi_p$ is expressed as
\be
\chi_p = \frac{N|{\mu}_{14}|^2}{\epsilon_0 \hbar \gamma_{41}} \chi
\ee
with
$$
chi = \frac{\tilde{\rho}_{41}\gamma_{41}}{|R_p|}
$$

\begin{figure*}
\begin{center}
\includegraphics[width=10cm,height=7cm]{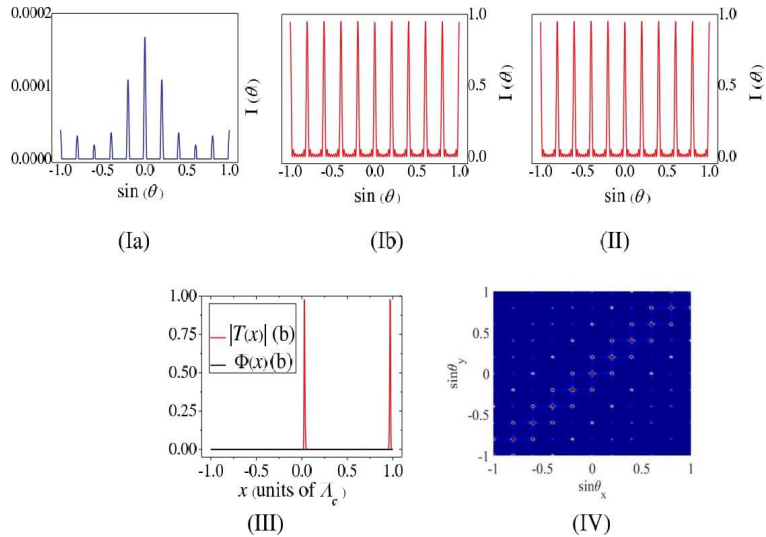}
\end{center}
\caption
{Resonant evolution of grating-spectra:
      I: Plot of Diffraction Intensity versus sin($\theta$) - (a)$R_m$$=$$0$, (b) $R_m$$=$$0.03$MHz. 
      II: Plot of Diffraction Intensity versus sin($\theta$) with same conditions as in I(b) considering only the nonlinear contribution ($f_{NL2}$ in the expression of $\frac{\rho_{41}}{R_p}$ (Eq. 6)).
      III: Plot of Transfer Function $T(x)$ (Red Colour) and phase $\Phi$ (Black Colour) with same conditions as in I(b).
      The other parameters: $R_p$$=$$0.01$MHz, $R_n$$=$$10$MHz, $R_c$$=$$8$MHz, $L$$=$$30$,
      and $\Delta_p$$=$$\Delta_n$$=$$\Delta_c$$=$$\Delta_m$$=$$0$.} 
\end{figure*}

In order to obtain the self-consistent equation for the probe
field propagating through the atomic medium, we consider the probe field
with a planar wavefront travelling along the $z$-direction, which is
represented as
$E_p$$=$$\frac{1}{2}\epsilon_pe^{i(\omega_pt - k_pz)}$$+$$c.c.$.
The amplitude factor $\epsilon_p$ is assumed to remain unchanged in the
transverse direction ($x$-direction) during the propagation of the wave
though the atomic medium. We define the propagation vector,
$k_p$$=$$\frac{2\pi}{\lambda_p}$; ${\lambda_p}$ being the wave length
of the probe field. Now, the Maxwell's equation for the probe
field is given in the following form [12]
\be
\frac{\partial{\epsilon_p}}{\partial{z}} =
                          i\frac{\pi}{\epsilon_0 \lambda_p} {\cal P}_p
         = i C \frac{\chi}{\lambda_p} {\epsilon_p}
\ee
where $C$$=$$\frac{{\pi}N|{\mu}_{14}|^2}
 {\epsilon_0{\hbar}{\gamma_{41}}}$ is a dimensionless constant [12,37].
For sake of the simplicity of the calculation, $C$ is chosen to be unity.
The product term $i\chi$ is redefined as ${\cal A}$$+$$i{\cal D}$ with
absorption ${\cal A}$$=$$\frac{\gamma_{41}}{R_p}A$
and dispersion ${\cal D}$$=$$\frac{\gamma_{41}}{R_p}D$, where
reduced absorption $A$$=$$-Im[\tilde{\rho}_{41}]$
and dispersion $D$$=$$Re[\tilde{\rho}_{41}]$.
Considering the dimensionless distance $\xi$$=$$z/{\lambda_p}$,
we recast the equation (9) as
\be
\frac{\partial{\epsilon_p}}{\partial{\xi}} = ({\cal A} + i{\cal D}) \epsilon_p,
\ee
which leads us to obtain the transmission function as
spatially-modulated-grating transfer function
\be
T({\xi},x) = T_o e^{[{\cal A}(x)+i{\cal D}(x)] \xi}
\ee
where $T_o$ is a constant.
For convenience, we consider the normalized transfer function
$f({\xi},x) = T({\xi},x)/T_o$, which plays the vital
role in producing the grating spectrum. Because,
all the information about coherent and incoherent atom-field interactions
is stored in this function. This is to mention here that $|f({\xi},x)|$
$i.e.$$e^{{\cal A}(x) \xi }$ is the amplitude part of the grating-transfer-function,
while the term $\Phi({\xi},x)$ ($=$${\cal D}(x)\xi$) evolves as the phase part of
the grating-transfer-function [4].
If the value of the amplitude transfer function becomes predominant
over the phase part of the transfer function, we obtain the intensity distribution of the grating induced mostly by the absorption. With the variation of the system-parameters, the intensity distribution pattern can be dramatically changed due to the significant contribution of
the phase part of the transfer function, which, in turn, leads to the formation of a phase grating.

We define the spatial width of the probe beam as
the product-term $M\Lambda_p$ where
$\Lambda_p$$=$${\lambda}_p/2$. $M$ is a non-zero positive integer,
which implies the allowed number of opto-atomic slits.
The amplitude of the probe field $\epsilon_p$ is taken to be uniform
over the width of the beam. On transmission through the slit-assembly,
the spatial modulation can be attributed to the probe
amplitude by introducing the function $G(x)$$=$$\epsilon_pf({\xi},x)$.
If $\theta$ be the angle of diffraction of the probe
field from the $z$-direction, then the Fourier transform of $G(x)$
gives rise to the resultant amplitude of the Fraunhofer diffraction pattern
as given bellow [38]
\be
A_P({\theta}) \propto
   \int_{-\infty}^{\infty} G(x)\,\, exp(-i\frac{2{\pi}}{\lambda_p} x sin{\theta}) dx
\ee
Introducing the dimensionless space-variable 
$x^{\prime}$$=$$\frac{x}{{\Lambda}_c}$, 
we newly define the parameter
$Q$$=$$\frac{{\Lambda}_{c}^{\prime}}{\Lambda_p}$,
Thus the intensity of the
diffraction pattern is expressed after the algebraic simplification
as follows [14,39]
\be
I(\theta) = I_o |F(\theta)|^2 \frac{\sin^2(M{\pi}Q\sin\theta)}
{M^2\sin^2({\pi}Q\sin\theta)}
\ee
where $I_o$ is the constant of proportionality including $|\epsilon_p|^2$
and $\Lambda_c^2$, and
\be
F(\theta) = \int_{-{1/2}}^{1/2} f(L) exp(-ip{\pi}Qx^{\prime}\sin\theta) dx^{\prime}
\ee
where $p$ is defined earlier,
The intensity of the $n$-th order diffraction maximum is specified
by using the grating equation $Q\sin\theta$$=$$n$.
The length $L$ indicates the dimensionless distance traversed by
the probe field through the active medium. We note that $|F(\theta)|^2$
implies the intensity distribution of the single slit diffraction.
In practice, for carefully chosen atomic transition to be probed,
$\Lambda^{\prime}_c$ may be of the order of $\Lambda_p$, because
it needs to adjust the value of $Fresnel$ $number$ [14] to be much less
than unity in the far-field diffraction regime.

\begin{figure*}
\begin{center}
\includegraphics[width=0.6\linewidth]{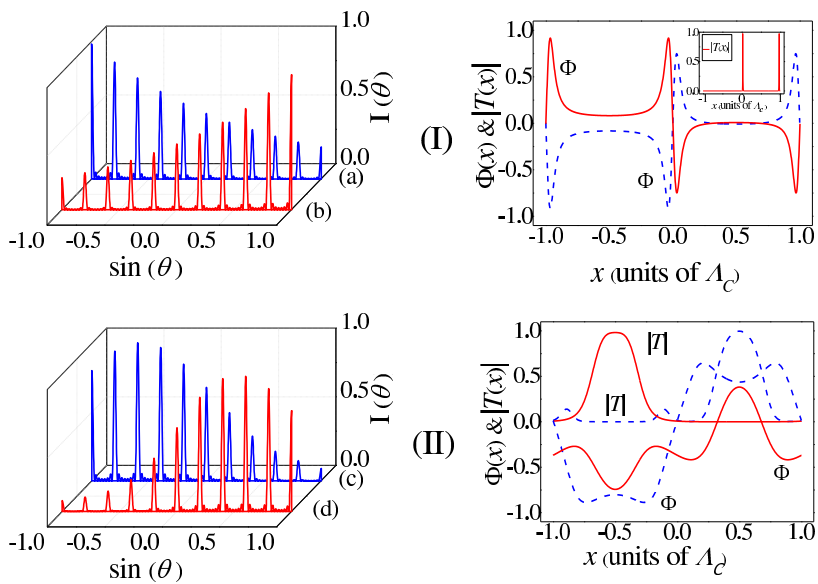}
\end{center}
\caption
{Near resonant evolution of grating-spectra:
      I: Left Panel - Plot of Diffraction Intensity versus sin($\theta$) - (a)$\Delta_p$$=$$\Delta_m$$=$$0.1$MHz, and (b)$\Delta_p$$=$$\Delta_m$$=$$-0.1$MHz; Right Panel: Corresponding plot of Transfer Function $T(x)$ (Inset) and phase $\Phi$.
      II: Left Panel - Plot of Diffraction Intensity versus sin($\theta$) - (c)$\Delta_p$$=$$\Delta_m$$=$$4$MHz, and (d)$\Delta_p$$=$$\Delta_m$$=$$12$MHz; Right Panel: Corresponding plot of Transfer Function $T(x)$ and phase $\Phi$. Same colours are used in (I) and (II) to indicate correspondence between the right and left panels.
      Other parameters: $R_p$$=$$0.01$MHz, $R_m$$=$$0.03$MHz, $R_n$$=$$10$MHz, $R_c$$=$$8$MHz, and $L$$=$$30$.} 
\end{figure*}
   
\section{Results and discussions}

We have computed numerically the Fraunhofer diffraction pattern of the probe
beam by using equations (6), (11), (13) and (14). First of all, we set the
values of $Q$$=$$5$ and $M$$=$$8$. The
value of $Q$ indicates that maximum number of peaks including the central
one is 11. The natural decay rates of the excited levels $|3>$ and $|4>$ are considered to be equal to $6$ MHz. For all the results presented here
we set the Rabi frequency of the probe laser as $R_p$ = 0.01 MHz and choose $\gamma_{31}$$=$$\gamma_{32}$$=$$\gamma_{42}$$=$$\gamma_{41}$ = 6 MHz. The values of the Rabi frequencies of the other two fields are taken as $R_n$$=$$10$MHz and $R_c$$=$$8$MHz in Fig.(2-4). At the condition of multi-photon resonance $i.e.$$\Delta_p$$=$$\Delta_c$$=$$\Delta_n$$=$$0$, for $L$$=$$30$, Fig.2 shows the variation of intensity $I_\theta$ of different peaks in the grating structure in the normalised scale and the corresponding plots of the amplitude part $|T(x)|$ and the phase part ${\Phi}(x)$ of $f({\xi},x)$. When the control field
specified by the Rabi frequency $R_m$ is switched off, the curve of
Fig.2(Ia) depicts the pattern of grating-spectrum [40] comprising of the complete
evolution of nine peaks accompanied by the significantly intense central maximum 
and incomplete appearance of the marginal peaks (of the fifth order) at near the both ends of sin($\theta$) axis. In this condition, the MWMIC-term does not contribute in shaping the 
grating-spectrum due to the absence of the spectral term $f_{NL2}$. 
In the presence of this nonlinear modulation
term with $R_m$$=$$0.03$MHz, the curve of Fig.2(Ib) shows the
appearance of almost equally enhanced peaks in the grating-spectrum. 
Such generation of the diffraction peaks in a supersymmetric fashion
occurs at the condition of exact atom-field resonance when the amplitude
part of the grating-transfer-function $f({\xi},x)$ only contributes in the process while
the phase part of $f({\xi},x)$ leaves no signature. This is remarkable
to note that similar feature of diffraction pattern (Fig.2(II)) mimics
exclusively for the grating-transfer-function generated by the nonlinear modulation term $f_{NL2}$ for the same parametric condition of Fig.2(Ib). For better apprehension of the physical condition $f({\xi},x)$, we have plotted Fig.2(III) where the red-line curve denotes the variation in  normalized amplitude transfer function ($|T(x)|$) and the dark-line signifies the null contribution of the phase function ${\Phi}(x)$.

On exploring such fascinating feature of uniform intensities for all the orders in the one-dimensional (1D) diffraction profile we proceed for the study of diffraction pattern in two dimensions aiming at potential application in atom lithography. Motivated by the scheme of sub-diffraction limited spots in quantum lithography as employed in ref.[41], we have presented the 2D density plot in Fig.2(IV) of the diffraction pattern obtained in the present model for the same set of values of parameters as chosen in Fig.2(Ib) shown for the 1D case. The expression of 2D intensity distribution $I(\theta_x, \theta_y)$ is given in Appendix A. Sharp features of the symmetrically arranged diffracted light spots make it possible to visualize periodic etching of the illuminated zone of the surface of the substrate used in lithography. Otherwise, on scanning the exposed surface of the substrate over the diffracted spots, optically written pattern will be generated, which is another aspect of obtaining high precision optical lithography.

To illustrate how the diffraction pattern evolves due to the resultant effect of two competitive components: amplitude and phase parts of $f({\xi},x)$ under the impact of nonlinear modulation, we plot  Fig.3 where the variation of intensity $I_\theta$ and the associated plots of the amplitude part $|T(x)|$ and the phase part ${\Phi}(x)$ of $f({\xi},x)$ are presented in the left and right panels respectively. When we switch on the knob of the nonlinear modulation i.e., fix $R_m$ at the same non-zero value (0.03 MHz) like Fig.2, at the parametric condition ($\Delta_c$$=$$\Delta_n$$=$$0$, $\Delta_p$$=$$\Delta_m$$=$$0.1$MHz, and $L$$=$$30$) the grating-spectrum (curve a in the left panel) of Fig.3(I) shows the monotonically decreasing nature of
the diffraction-peak intensity from the extreme left end to the extreme right. This
nature of intensity variation is reversed (curve $b$ in the left panel of Fig.3(I)) with the negative detuning $\Delta_p$$=$$\Delta_m$$=$$-0.1$MHz when the other parameters remain same. We note that the figure, as shown by the curves $a$ and $b$ together, exhibit evolution of gradually increasing / decreasing peak intensity from one end to another thereby inducing asymmetry in peak pattern.
For these two conditions, it is prominent in the right panel of Fig.3(I) that the nature of the amplitude part $|T(x)|$ does not change for the cases of the curves of $a$ and $b$, while the phase part ${\Phi}(x)$ varies in counter opposite way. We note that the curves are resulted from the combined effect of $|T(x)|$ and ${\Phi}(x)$.
When only the rate of detuning is increased to some higher value
like $\Delta_p$$=$$\Delta_m$$=$$4$MHz for other fixed parameters, we obtain the enhanced peaks at the left half of sin$\theta$ axis as displayed by the curve $c$ in left panel of Fig.3(II). The peaks in the right half are less significant in comparison the peaks of the left
half. For more higher values of the detuning like
like $\Delta_p$$=$$\Delta_m$$=$$12$MHz, the feature of peak enhancement shows an opposite variation i.e., the peaks in the right half become prominent in comparison to the peaks arising in the
left half as shown by the curve $d$ in the left panel of Fig.3(II) for the same values of the other parameters. The nature of variation of $|T(x)|$ and ${\Phi}(x)$ in the right panel of Fig.3(II), does not predict the exact dominance of any of these two functions in the formation of such diffraction structure for each of the curves $c$ and $d$.

\begin{figure*}
\begin{center}
\includegraphics[width=0.5\linewidth]{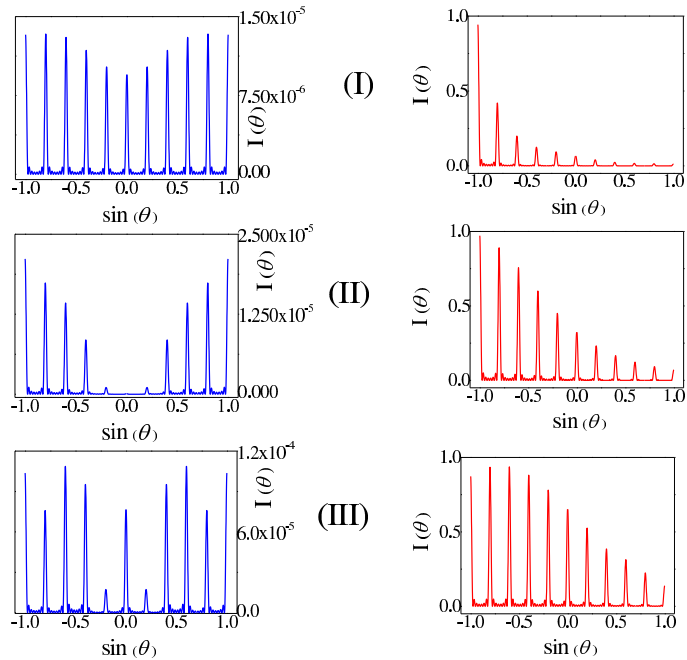}
\end{center}
\caption
{Sample length dependence of grating-spectra at near resonance:
    (I) $\Delta_p$$=$$\Delta_m$$=$$1.5$MHz and $L$$=$$60$,
    (II) $\Delta_p$$=$$\Delta_m$$=$$3.5$MHz and $L$$=$$70$,
    (III) $\Delta_p$$=$$\Delta_m$$=$$5.2$MHz and $L$$=$$70$, 
    with the other parameters: $R_p$$=$$0.01$MHz, $R_n$$=$$10$MHz,
    $R_c$$=$$8$MHz and $\Delta_n$$=$$\Delta_c$$=$$0$. All the figures in the left panels are drawn for $R_m$$=$$0$ and the figures in the right planes for $R_m$$=$$0.02$MHz.} 
\end{figure*} 
In order to visualize the effect of the length of the active medium on the
diffraction pattern at the detuned condition of the probe, we show the figures in the left and right panels of Fig.4(I,II,II) in the absence and presence of the nonlinear modulation terms respectively for the fixed values of the parameters like
$\Delta_c$$=$$\Delta_n$$=$$0$. At the switched-off condition of the laser responsible for $R_m$ the curve shown in the left panel of Fig.4(I) is dealt with the parameters:  $\Delta_p$$=$$1.5$MHz, and $L$$=$$60$. It is observed that 
the intensities of the peaks of non-zero order number get regularly enhanced with the increase in the order number, as we shift from the central peak of minimum intensity. But
the overall intensity of the peaks become less significant. This is because
of the predominant role of the phase part of $f({\xi},x)$ in the absence of
$f_{NL2}$. In the presence of the control field with $R_m$$=$$0.02$MHz, the curve given at the right panel of Fig.4(I)
gives rise to the evolution of asymmetric peaks with significantly reduced 
intensities of the peaks in the right half, which is due to the role of MWMIC regulating the
peak-pattern. With a small increase of the value of the probe detuning
($\Delta_p$$=$$3.5$MHz) for the active medium with increasing length $L = 70$ in the absence of the $R_m$, the curve shown in the left panel of Fig.4(II)
shows that the central peak almost vanishes with the
occurrence of the enhanced higher order peaks on both sides of the grating spectrum. But, in the curve given at the right panel of Fig.4(II)
($R_m$$=$$0.02$MHz), we see that the nonlinear modulation of peak intensities
by the MWMIC-effect increases the intensities of the peak-pattern towards the right end from the extreme left peak.
For further increase in the detuning like $\Delta_p = \Delta_m = 5.2$ MHz keeping $R_m$ field switched off and the other parameters same, it is observed that the curve exhibited in the left panel without MWMIC-effect becomes significantly modified in presence of MWMIC-effect as displayed by the curve in the right panel of Fig.4(III). This is to mention here that the modulation appearing in the peak intensities at the left half of the curve shown in the right panel is significantly different when compared to that attained in the other two curves as shown in the right panel of Fig.4(I,II). Thus, we can infer that the term $f_{NL2}$ plays the key role to modify the intensity of the diffraction peak.

It is well known that, for the evolution of $i$-th peak in the grating spectrum, $I(\theta_i)$/$I_0$ measures the diffraction efficiency (D.E.) [8] for that particular diffraction peak. In this context, we have drawn the ratio of DE $i.e.$ $I(\theta_i)$/ $I(\theta_j)$ ($i>j$)in Fig.5 for $i$=4(sin$(\theta)$ = 0.8) and $j$=1(sin$(\theta)$ = 0.2). Fig.5(I,II,III,IV) exhibit the mentioned ratio of D.E. with the increase of the values of the
controlling parameters like the Rabi frequencies: $R_n$ (Fig.5(I)),
$R_c$ (Fig.5(II)), the detuning parameter $\Delta_p$$=$$\Delta_m$
$=$$\Delta$ (Fig.5(III)) and the length $L$ of the active medium (Fig.5(IV)) 
for fixed values of the parameters like
$\Delta_c$$=$$\Delta_n$$=$$0$, $R_p$$=$$0.01$MHz and $R_m$$=$$0.03$MHz.
The value of $L$ is set as $30$ for Fig.5(I-III).
As depicted by Fig.5(I) ($\Delta$$=$$0$, $R_c$$=$$8$MHz), the intensity ratio rapidly increases up to $R_n$ $=$ $10$MHz and then slowly comes into saturation for larger values of $R_n$. 
If we plot the intensity ratio versus the Rabi frequency $R_c$, we obtain the curve as presented by Fig.5(II)($\Delta$$=$$0$, $R_n$$=$$5$MHz). It is observed that with the increase in the value of $R_c$, the ratio decreases. This is surprising that one can obtain double peak like structure in the variation of the ratio with the probe-detuning as depicted by the plot of Fig.5(III) with $R_n$$=$$10$MHz and $R_c$$=$$8$MHz. This implies that higher order peak intensity is significantly enhanced with respect to that of the first order. In between the two peaks the ratio is very small. This feature can be attributed to the negative impact of the interplay between the amplitude and
phase parts of the transfer function $f({\xi},x)$ on the formation of the
diffraction pattern. In Fig.5(IV) ($\Delta$$=$$5.2$, $R_n$$=$$10$ and
$R_c$$=$$8$), we have shown the interesting feature for the variation of the intensity ratio with the increase of the length of the active medium. This implies the appearance of the diffraction peaks with equal intensity with the increase of sample length. Thus, it can be concluded that proper choice of the values of the controlling parameters involved in the system leads us to obtain sufficiently intense higher order diffraction peak intensities at resonant as well as off-resonant conditions of the probe field.

\begin{figure*}
\begin{center}
\includegraphics[width=0.4\linewidth]{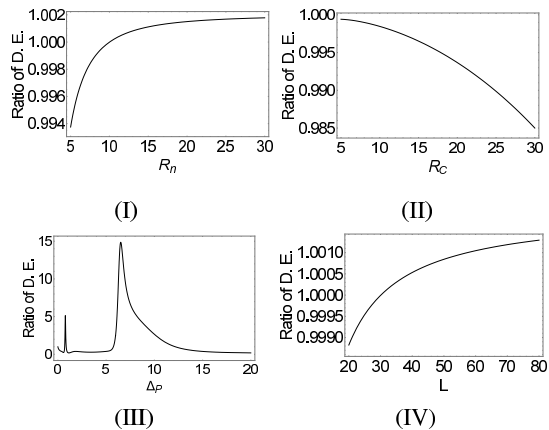}
\end{center}
\caption
{Ratio of diffraction efficiencies (D.E.)to show Rabi frequency, probe-detuning and sample length dependence for $i = 4$ for the fourth order peak at sin$(\theta)$ = 0.8 and $j = 1$ for the first order peak at sin$(\theta)$ = 0.2 in Fig. 2(Ib). All the parameters are same as in Fig. 2(Ib) except the variable used.} 
\end{figure*} 

\section{Conclusion}

We have studied nonlinear modulation effect on the diffraction pattern of EIG as a result of spatially modulated four-wave-mixing induced coherence effect in a four-level close-loop interaction system. In order to enhance the applicability of the proposed scheme in practice, Rubidium 87$D_1$ transitions are taken into account to obtain the grating pattern. 
We note that the amplitude part of the grating-transfer-function plays the vital role in forming supersymmetric and significantly enhanced diffraction peaks at the condition of exact atom-field resonance. It has been described how peak-asymmetry results in the system at the detuned condition of the probe field. This is interesting to observe that higher order peaks can be much more enhanced than that of the lower order peaks due to nonlinear peak modulation effect. We have shown that, with the variation of the length of the active medium, the diffraction pattern obtained at a particular parametric condition can be changed abruptly. The method of achieving controllable peak-pattern may be an useful technique for generation, control and detection of nonlinear light originated by the spatially modulated four-wave-mixing process. The variation in peak-pattern suggests that the present model may be applied in case of high-precision atomic lithography. In this context, a density plot of the supersymmetric diffraction peaks in two dimensions is given to realize the efficacy of the model, which seems to be appropriate for making periodic etching optically the surface of the active medium to be taken as the substrate in lithography.

\section{Appendix}

\noindent Considering the probe field to be a plane wave, we obtain the resultant
intensity of the far-field diffraction pattern by making the Fourier
transform of $f({\xi},x,y)$ for a sample-length ${\xi}$ $=$ $L$, which
is given by introducing the parameters:
$Q_x$ $=$ $\frac{{\Lambda}_{cx}}{\Lambda_p}$ and
$Q_y$ $=$ $\frac{{\Lambda}_{cy}}{\Lambda_p}$
with $\Lambda_p$ $=$ ${\lambda}_p/2$,
\renewcommand{\theequation}{A.1}
\begin{widetext}
\be
I(\theta_x,\theta_y) = I_0 |F(\theta_x,\theta_y)|^2
\frac{\sin^2({\pi}M_xQ_x\sin\theta_x)}{M_x^2\sin^2({\pi}Q_x\sin\theta_x)}
\frac{\sin^2({\pi}M_yQ_y\sin\theta_y)}{M_y^2\sin^2({\pi}Q_y\sin\theta_y)}
\end{equation}
\end{widetext}

where $I_0$ is the constant of proportionality including $|\epsilon_p|^2$,
$\Lambda_{cx}^2$ and $\Lambda_{cy}^2$. Introducing the dimensionless
space-variables: $x^{\prime}$ $=$ $\frac{x}{{\Lambda}_{cx}}$ and
$y^{\prime}$ $=$ $\frac{y}{{\Lambda}_{cy}}$, we express
\renewcommand{\theequation}{A.2}
\begin{widetext}
\be
F(\theta_x,\theta_y) =
\int_{-1/2}^{1/2} dx^{\prime} \int_{-1/2}^{1/2} dy^{\prime}
f(L,x^{\prime},y^{\prime})exp(-i{\pi}Q_xx^{\prime}\sin\theta_x)
exp(-i{\pi}Q_yy^{\prime}\sin\theta_y)
\ee
\end{widetext}

\noindent where the spatial width of the probe beam is defined
along $x$- and $y$- directions by the product-terms ($M_x\Lambda_p$)
and ($M_y\Lambda_p$) respectively. $M_x$ and $M_y$ are
nonzero positive integers, which imply the allowed number
of opto-atomic slits in the $xy$-plane as required for a
cross-grating. The $(m,n)$-order diffraction peak is
determined by the grating equations:
$Q_x\sin\theta_x$ $=$ $m$ and $Q_y\sin\theta_y$ $=$ $n$.

\section{Acknowledgments}

PP thankfully acknowledge the research centre in PKC for support. BKD likes to acknowledge the tenure of his service in J. K. College, Purulia because he felt motivation to do research in this direction during this period.

\section{Corresponding author}

Correspondence to ppcontai@gmail.com


\begin{thebibliography}{99}
\bibitem{39}
In ref.[14], the expression (17) should contain the term $|F(\theta)|^2$ instead of $F(\theta)$. This is a typographical mistake.
\bibitem{40}
The word ‘spectrum’ is generally used for the distribution of intensities over wavelengths. Here, ‘grating-spectrum’ implies the appearance of the angular distribution of diffraction peaks over the variation of spatial phase at the output of the grating.
\end{thebibliography}
\end{document}